\documentclass[twocolumn,showpacs,prl]{revtex4}
\input{epsf}

\begin{document}

\title{\bf Supersymmetric manipulation of quasienergy states. Application
to the Berry phase}

\author{Boris F. Samsonov$^1$\footnote{On leave from the Physics Department, Tomsk State
University, Russia},
M.~L.~Glasser$^2$ and
L.~M.~Nieto$^1$}
\affiliation{$^1$Departamento de F\'{\i}sica Te\'orica, Universidad de
Valladolid, 47005 Valladolid, Spain
\\
$^2$Department of Physics and Center for Quantum Device Technology,
Clarkson University, Potsdam NY 13699-5820 (USA)}

\date{\today}

\begin{abstract}
Time-dependent supersymmetry allows one to delete quasienergy
levels for time-periodic Hamiltonians and to create new ones. We
illustrate this by examining an exactly solvable model related to
the simple harmonic oscillator with a time-varying frequency. For
an interesting nonharmonic example we present the change of the
Berry phase due to a supersymmetry transformation.

\end{abstract}


\maketitle
%%%%%%%%%%%%%%%%%%%%%%%%%%%%%%%%%%%%%%%%%%%%%%%

One may say without exaggeration that non-relativistic quantum
mechanics is essential to understanding the basic laws of nature.
 Thus, the Schr\"odinger equation is one of the most
fundamental constructs of  modern natural science and
over the past century, as evidenced by a vast literature, great progress
 has been made in developing
various consequences of the {\it time-independent} Schr\"odinger
equation describing stationary processes. Key recent advances concern
 the manipulation of bound states \cite{Zakhariev}.
 In the real world, however, nearly everything changes with time, and
  stationary processes are exceptional (perhaps even impossible), but
 there is ample evidence
 for believing that there will soon be similar advances in
the
 study of non-stationary processes
and the consequences of the {\it time-dependent} Schr\"odinger equation.
The end of the last century saw some progress in this area:
an extensive investigation into the properties of  {\it time periodic}
 Hamiltonians, their
dynamical invariants \cite{MMK} and symmetries \cite{Alon},
dynamical pulsed coherent states \cite{GV}, Floquet quanta
\cite{Mielnik} and  application in optics \cite{WSG}.
An area that
 is now attracting interest is  calculating the Berry, or
geometric,
  phase for time dependent Hamiltonians, although so far the only system treated in
   detail is the,
exactly solvable, harmonic oscillator with time dependent mass and/or frequency
 \cite{GeChild}.

 It has
long been known how to construct new exactly solvable time-independent Hamiltonians from an
initial
 one by means of the Darboux, or supersymmetry,
transformation (see e.g. \cite{books,BSRev}), and the
 procedure has recently been generalized to include
 time dependence \cite{BSRev,BSPLA}. Although
stationary state energies do not exist for a time-dependent Hamiltonian,
 if the time dependence is periodic,
{\it quasienergies} play a similar role \cite{Zeldovich,Per}.
 In this letter, based on the simple
 harmonic oscillator model, having a time-periodic  frequency, we
show that this analogy  is much deeper than previously recognized
and that it can be extended to the level of supersymmetry. In
particular, we show that  processes such as the creation and
annihilation of quasienergies may be effected easily by
time-dependent supersymmetry (or Darboux) transformations. This
also opens the door for performing more complex operations on
quasienergy states by applying  integral transformations which are
essentially  inverses to the differential supersymmetry
transformations \cite{BSRev}, \cite{Zakhariev,books}.

Since exact eigenstates are known for these new time-dependent
Hamiltonians their remarkable properties,
such as
 the the Berry phase, can be investigated in detail.
  We demonstrate here that the Berry  phase,
 for systems generated from our model harmonic oscillator
Hamiltonian, simply acquires an additive correction under a
Darboux transformation.

We begin with the Schr\"odinger equation for the Hamiltonian
\begin{equation} \label{h0}
h_0=-\partial _x^2+\omega ^2(t)x^2 \,,\quad
\omega (t+T)=\omega (t)
\end{equation}
where $\partial_x\equiv \partial /\partial_x$ and
 appropriate units have been chosen. A complete set of solutions
normalized to unity
on the real line is  \cite{LR} \cite{Man}
\begin{equation}\label{psin}
\psi_n(x,t)=
\left( \overline\varepsilon/\varepsilon\right) ^{n/2+1/4}
\exp \left( i\dot \gamma z^2\right)f_n(x,t)
\end{equation}
where
\begin{equation}\label{fn}
f_n(x,t)= N_n\gamma ^{-1/4}
 H_n(z )  e^{ -\,z^2/2} \,,\
 z(x,t)=x/\sqrt{8\gamma(t)}\,,
\end{equation}
Here
$N_n=(2^{n+1}n!\sqrt{2\pi})^{-1/2}$,
$ \gamma =\varepsilon \overline \varepsilon  $,
$H_n(z)$ is a Hermite polynomial and $\varepsilon(t)$,
$\overline \varepsilon(t)$ are two linearly independent solutions
of the classical equation of motion for the harmonic oscillator
 \begin{equation}\label{eps}
\ddot \varepsilon (t)+4\omega ^2(t)\varepsilon (t)=0\,.
 \end{equation}
 A dot over a symbol represents the derivative with respect to
 time and
 we choose these solutions so that their Wronskian
 $\dot\epsilon(t)\bar\epsilon(t)
 -\epsilon(t)\dot{\bar\epsilon}(t)=\frac i2$.
 In general, the solutions of  (\ref{eps}) are of three types: (i) both $\varepsilon$ and $\overline\varepsilon$
 are stable, (ii) they are both unstable, (iii) one is stable the other
is
 unstable.  It is known \cite{Per} that only in the first case the
functions
 (\ref{psin}) have the property
  \begin{equation}\label{epsT}
  \psi_n(x,t+T)=e^{-iE_nT} \psi_n(x,t)\,,
  \end{equation}
 where $E_n=(n+\textstyle{\frac 12})\delta$ is called the {\it quasienergy}
 \cite{Zeldovich}
 and
 $\delta $ is defined by the equation
 \begin{equation}\label{delta}
\varepsilon (t+T)=\varepsilon (t)e^{i\delta T}  \,.
  \end{equation}
According to Floquet's theorem a (complex)
solution $\varepsilon (t)$ to equation (\ref{eps}) satisfying
(\ref{delta}) can always be found in  case (i). We
choose the complex conjugate of $\varepsilon$ to be the second, linearly
independent, solution to (\ref{eps}). Here and below
 a bar over a quantity denotes its complex conjugate and we note that
$\gamma (t)$ is real. It is clear from (\ref{eps}) that
quasienergies $E_n$ are defined modulo $2\pi$ and if they are
comensurable with $\frac{2\pi}{T}$, the functions $\psi_n$ are
periodic.

A further useful property of the functions (\ref{psin})  is that there
exist ladder operators
 \begin{equation}\label{a}
 a=\varepsilon \partial _x-i\dot \varepsilon x/2, \quad
 a^+=-\bar \varepsilon  \partial _x+i {\dot{\bar \varepsilon} }x/2
 \end{equation}
 such that
 \begin{equation}\label{lad}
a\psi_n=\frac{\sqrt n}{2}\,\psi_{n-1}\,,\quad
a^+\psi_n=\frac{\sqrt {n+1}}{2}\,\psi_{n+1}\,.
 \end{equation}
These functions are eigenfunctions of the symmetry operator
$g=(aa^++a^+a)/2$, $g\psi_n=g_n\psi_n$ with $g_n= \frac 18(2n+1)$.

Next we shall apply the approach presented in \cite{BSPLA} for
constructing new exactly solvable time-dependent Hamiltonians having
the same quasienergy spectrum, with the possible exception of a few
levels, as a given one. The procedure is simply the time-dependent
generalization of the usual supersymmetry
construction \cite{books}. A new hamiltonian $h_1$ is defined by means
of a  nodeless solution $u(x,t)$,
called the {\it transformation function}, to the initial
Schr\"odinger equation, $i\partial_t u(x,t)=h_0u(x,t)$,
subject to the
additional condition
$\partial_x^3 \left( \ln u/\overline u\right) =0 $,
which guarantees that  $h_1$ has the form
 \begin{equation}\label{h1}
 h_1=h_0-\partial_x^2\ln |u(x,t)|^2\,.
 \end{equation}
 Solutions of the corresponding Schr\"odinger equation
 are obtained by applying
the differential  operator \begin{equation}\label{L}
L=L_1(t)[-\partial_x+\partial \ln u(x,t)/\partial x]
\end{equation}
to $\psi_n$:
\begin{equation}\label{fin}
\varphi_n(x,t)=M_n^{-1/2}L\psi_n(x,t)
\end{equation}
where the time-independent factor $M_n^{-1/2}$
guarantees that the functions (\ref{fin}) are normalized
to unity.
The operator $L$ is defined in terms of the
same function $u(x,t)$ as in (9)
and
\begin{equation}  \label{L1}
L_1\left( t\right) =\exp \left[ 2\int
\mathop{\rm Im}\left( \partial_x^2\ln u\right) dt\right]  \,.
\end{equation}
The operator $L^+=L_1(t)[\partial_x+\partial\ln\overline u(x,t)/\partial x]$,
 adjoint to $L$,
realizes the transformation in the opposite direction, from solutions
of the Schr\"odinger equation with  Hamiltonian $h_1$ to
those of the Hamiltonian $h_0$. Thus, the superposition $L^+L$
is a symmetry operator for the initial Schr\"odinger equation and the
function $u(x,t)$ is an eigenfunction of this operator:
$L^+Lu(x,t)=g_uu(x,t)$. The normalization factor in (\ref{fin})
is equal to its mean value
$M_n=\langle \psi_n|L^+L|\psi_n\rangle$.

Among the functions (\ref{psin}) only $\psi_0$ is nodeless and
suitable for using as a transformation function; it produces only a shift of the Hamiltonian by an
$x$-independent value. This is simply a manifestation of the
well-known shape-invariance property \cite{books} in the
time-dependent case.
Any other function $u_k(x,t)=\psi_k(x,t)$, $k>0$ taken as the
transformation
function will produce a potential with $k$ poles corresponding
to the zeros of $\psi_k(x,t)$ which clearly has no  physical meaning
if the variable $x$  runs over the whole real line.
The transformed Hamiltonian $h_1$ can be taken as the initial
one for the next transformation step and if this is realized
with the transformation function
$\widetilde u_{k+1}(x,t)=Lu_{k+1}(x,t)$ all the poles are removed
and the resulting Hamiltonian $h_2^{(k)}=h_0-A_2^{(k)}(x,t)$ is physically admissible.
For the potential difference one gets
\begin{equation}\label{eq103}
A^{(k)}_2(x,t)= \frac {1}{4\gamma }\left[ \frac {J''_k( z)}{J_k(
z)} - \left( \frac {J'_k( z)}{J_k( z)}  \right) ^2 -2 \right] \,.
\end{equation}
Here
$$J_k\left( z\right) =\sum_{j=0}^k\frac{k! }{2^{j}j!}
H_j^2\left(\textstyle{ z}\right)
$$
and for the first three of these functions  one has the  simple
expressions:
$$J_0\left( z\right) =1,\quad J_1\left( z\right) =2z^2+1,\quad J_2\left(
z\right) =4z^4+3\,. $$
The Hamiltonian $h_2$ has the same system of quasienergies as
$h_0$ except for those corresponding to $n=k$ and $n=k+1$ which
are now deleted.
We can repeat this process without any
restrictions and get a Hamiltonian having an initial quasienergy
spectrum with any number of lacunae composed of two adjacent
levels.
 The potential $V_2^{(k)}=\omega^2(t)x^2-A_2^{(k)}$
looks like a harmonic oscillator potential, at the bottom
of which there are $k$ additional minima. The behavior of the $k=2$ case,
which is typical, is sketched
in Fig.~1 together with the harmonic oscillator potential.
%%%%%%%%%%%%%%%%%%%%%%%%%%%%%%%%%%%%%%%%%%%%%%%%%%%%%%%%%%%%%%%%%%%%%%%%%%%%%%%
%%%%%%%%%

 \begin{figure}
\epsfysize=3.5cm \epsffile{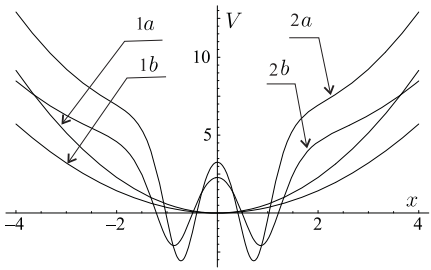} \caption{\small Potentials
$V_2^{(2)}(x,t)$ at $t=0$ (curve 2a) and at $t=T/2$ (curve 2b)
together with the harmonic oscillator potential (curves 1a and 1b
respectively). } \label{fig1}
\end{figure}

The opposite process, the creation of new quasienergy levels,
is also possible. For this purpose we need {\it unphysical}
solutions of the initial Schr\"odinger equation,
which do not belong to the Hilbert space and have growing
asymptotic behavior as $|x|\to \infty$ at any fixed  moment of time. It
is not difficult to see that the functions
\begin{equation} \label{uk}
u _k(x,t) = \gamma ^{-1/4}
\left(\epsilon /\overline \epsilon \right) ^{k/2+1/4}
H_k\left( iz\right)
e^{ (2i\dot \gamma +1/2)z^2}
\end{equation}
have this property.
 For any even value  $k=2l$ these are nodeless and suitable
 for using as transformation functions. They produce the
 new Hamiltonians $h_1^{(2l)}=-\partial_x^2+V_1^{(2l)}(x,t)$,
 $l=0,1,\ldots $
 with potentials
\begin{equation}
V_1^{(2l)}(x,t)=\omega ^2(t)x^2-A_1^{(2l)}(x,t), \ \ L_1(t)=\sqrt{
\gamma(t)} \, ,
\label{eq100}
\end{equation}
where
\begin{equation}
A_{1}^{(2l)}=\frac {1}{4\gamma }\left[ 1+4l(2l-1)\frac {q_{2l-2(
z)}}{q_{2l( z)}}
-8l^2\left(  \frac{q_{2l-1}( z)}{q_{2l}( z)} \right) ^2\right] \!,
\label{eq101}
\end{equation}
and
$q_k(z)=(-i)^k2^{-k/2}H_k(iz)$.
Using the recursion  relation for Hermite polynomials one finds
$$
q_0(z)=1,\ q_1(z)=\sqrt{2}z,\ q_{k+1}(z)=\sqrt{2}zq_k(z)+kq_{k-1}(z)\,.
 $$
The fact that a new quasienergy level is created by this process
follows from the property that the function
$v=1/(L_1\overline u_k)$
is a square integrable solution of the transformed
Schr\"odinger equation. It is easy to see that it corresponds to
the quasienergy $E=-\delta (k+\frac 12)$, which, in general, is
different from all the other quasienergies $E_n=\delta (n+\frac
12)$, $n=0,1,\ldots $. The case $k=0$  reproduces the  harmonic
oscillator Hamiltonian shifted by an $x$-independent quantity.
The
first nontrivial case corresponds to $k=2$. We display a typical
potential, $V_1^{(2)}(x,t)$, in Fig.~2.
%%%%%%%%%%%%%%%%%%%%%%%%%%%%%%%%%%%%%%%%%%%%%%%%%%%%%%%%%%%%%%%%%%%%%%%%%%%%%%%

\begin{figure}
\epsfysize=3.5cm
\epsffile{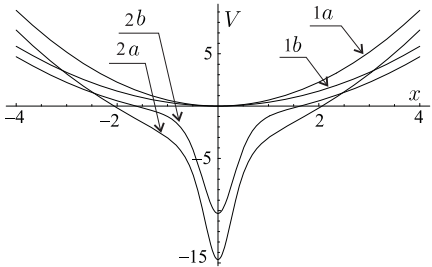}
 \caption{\small Potential
 $V_1^{(2)}(x,t)$ at $t=0$
(curve 2a) and at $t=T/2$ (curve 2b) together with the harmonic
oscillator potential (curves 1a and 1b respectively).
 }
 \label{fig2}
\end{figure}

After applying the operator (\ref{L}) to the functions
(\ref{psin}) one gets solutions of the transformed equation
corresponding to the quasienergies $E_n$.
For the case $k=2$ they
can easily be expressed in terms of
$\psi_n(x,t)$:
\begin{equation} \label{fin2}
\begin{array}{l}
\varphi_n(x,t)=\frac{1}{\sqrt{n+3}}
\left[
 \sqrt{n+1}(\epsilon /\bar \epsilon )^{1/2}\psi_{n+1}(x,t)\right.
 \\
 \left.
 ~~~~~~~~~
 +
 \frac{\sqrt 2\, z}{z^2+1/2}\psi_n(x,t)
\right]  \,.
\end{array}
\end{equation}
The normalization factor  here is calculated by noting that
the symmetry operator $L^+L$ is simply the shifted  $g$ operator,
$L^+L=g+5/8$.

We now turn to the transformation of the Berry phase and
consider for simplicity  the first nontrivial case $k=2$.
In these calculations we are using the standard approach (see
 e.g. \cite{AA}).
 The  phase change
 of a solution of the Schr\"odinger equation $\psi(x,t)$
 during time $T$ is given
 in terms of the mean energy value
 \begin{equation}
  \chi=-\int _0^T\langle \psi |i\dot \psi\rangle dt
 \end{equation}
  and the Berry phase $\beta$ is obtained from this overall
  phase change,
 to be extracted from (2) for the harmonic oscillator
 potential and from (17) for transformed potentials,
 by subtracting the dynamical component.

  For the functions in (\ref{psin}) one gets
 \begin{equation} \label{aven}
\langle \psi_n |i\dot \psi_n\rangle =
\frac{i}4(2n+1)\frac{d}{dt}{\ln}(\bar\epsilon /\epsilon)-
\frac 18 \langle \psi_n |x^2|\psi_n\rangle
\frac {d^2}{dt^2}{\ln}\gamma
 \end{equation}
 where the mean value of the square coordinate is
 $\langle \psi_n |x^2|\psi_n\rangle =4\gamma (2n+1)$.

  The structure of the functions $\varphi_n(x,t)$ is similar
  to those in (\ref{psin}) and therefore the  average energy is given by
  (\ref{aven}) with the replacement
  $\psi_n\to\varphi_n$.
To calculate
 the mean value of the square coordinate for the functions
(\ref{fin2}) we use the property $L^+L=g+5/8$ and the fact that
the functions (\ref{psin}) are eigenfunctions for $g$ to get
$$
\langle \varphi_n |x^2|\varphi_n\rangle =
\langle \psi_n |x^2|\psi_n\rangle +
\frac{4\sqrt \gamma}{\sqrt{n+3}}\langle \psi_n |x|\varphi_n\rangle \,.
$$
The first integral here is standard and the second has the  expression
$\langle \psi_n |x|\varphi_n\rangle =\sqrt\gamma \, [n+3-I_n(\frac 12)],$
where
$$
I_n(a)=\int_{-\infty}^\infty \frac{H_n^2(x)}{x^2+a}\,e^{-x^2}dx
$$
 which obeys the recursion relation
$$
I_n(a)=-2I_{n-1}(a)+4(n-1)^2I_{n-2}(a)-4a I'_{n-1}(a),
$$
which follows directly from that for
the Hermite polynomials. The initial
values are
$I_0(a)=\frac{\pi}{\sqrt a}e^a\mbox{erfc}\sqrt a$ and
$I_1(a)=4\sqrt \pi-4aI_0(a)$. Therefore, the mean
energy in the states (\ref{fin2}) is given by
 \begin{equation} \label{avenfi}
\langle \varphi_n |i\dot \varphi_n\rangle =
\langle \psi_n |i\dot \psi_n\rangle -
\left(1-\frac{I_n(\frac 12)}{n+3}\right)\,\gamma
\frac{d^2}{dt^2}{\ln}\gamma \,.
 \end{equation}

After being integrated over a period $T$ the first term
in (\ref{aven}) gives us the overall phase change
both for the initial states and for the transformed ones
(see  (\ref{avenfi}));
 the Berry phase is determined only by the second term.
Thus, the Berry phases $\beta_n^0$ for all states $\psi_n(x,t)$
are determined by the Berry phase for the ground state:
$\beta_n^0=(2n+1)\beta_0^0$,
$$
\beta_0^0=\frac 12\int_0^T\gamma\frac{d^2}{dt^2 }{\ln}\gamma dt=
-\frac 12\int_0^T\frac{\dot\gamma^2}{\gamma}\,dt \,.
$$
For the transformed Berry phase from (\ref{avenfi}) one gets
$\beta_n^1=\beta_n^0+2[1-\frac{I_n(1/2)}{n+3}]\beta_0^0$.
 Hence,
 once the quantity $\beta _0^0$ is known,
 we can easily calculate the Berry phase
 both for the harmonic oscillator and
 for the Hamiltonian $h_1^{(2)}$.

 As a  numerical illustration
 we have chosen a model for which analytic
 solutions of equation (\ref{eps})
 are available:
 $\omega (t)= \sqrt{\omega_0^2-\frac 12\wp (t+\omega_i)}$.
 Here $\omega_0$ is a parameter of the model along with the
 real ($\omega_r=T$) and imaginary ($\omega_i$) periods
 of the Weierstrass $\wp$ function. If we eliminate
 the parameter $\omega_0$ in favor of $d$ given by
 $\wp(d)=-4\omega_0^2$,
 the solutions of the equation  (\ref{eps})
 have the form \cite{WitVat}
 $$
  \varepsilon (t)=\frac{\sigma (t+\omega_i+d)}{\sigma (t+\omega_i)}
  e^{-t\zeta (d)}\,.
 $$
Here $\sigma $ and $\zeta $ are (non-elliptic) Weierstrass
functions. Figures 1 and 2 are plotted with
$\omega_r=-i\omega_i=2$ and $\omega_0\cong 0.5978$. We have found
also the  value $\beta_0^0=-0.0149$.

  We make one further comment concerning the time-dependent supersymmetry
 underlying our approach. The transformation operators are related to
 eigenfunctions (not necessarily ``physical") of the symmetry
 operator $g=\frac 12 (a^+a+aa^+)$, $gu_k=g_uu_k$,
 $g_u=5/8$,
 which accounts for the factorization $L^+L=g-g_u$.
 When the order of transformation operators is interchanged,
 one gets a symmetry operator for the transformed Schr\"odinger
 equation $\widetilde g$, $LL^+=\widetilde g-g_u$. The operators
 $g$ and $\widetilde g$ are supersymmetric partners from which
 a matrix operator
 ${\cal G}=\mbox{diag}(g, \widetilde g)$ can be constructed. It acts in
the space
 spanned by the basis vectors $\Psi_n^{(1)}=(\psi_n,0)^t$ and
 $\Psi_n^{(2)}=(0,\varphi_n)^t$ where the superscript ``$t$" stands
 for transposition and the functions $\psi_n$ and $\varphi_n$
 are given in (\ref{psin}) and (\ref{fin2}) respectively. So,
 just as in the usual supersymmetric approach, one has
 a two-fold degenerate spectrum except for the ground state
 level which is non-degenerate. This means that we have constructed here
 a model with unbroken supersymmetry.
 Finally we note that
 supercharge operators, closing superalgebra, can be constructed
 as usual with the help of the transformation operators $L$ and
 $L^+$.

In summary, in this note we have constructed, in principle, an
infinite number of time-periodic Hamiltonians having almost
unlimited complexity, for which the Berry phase can be determined
explicitly and  have illustrated the procedure for a Hamiltonian
whose time dependence is given by the square root of an elliptic
function. This vastly extends the set of previously known cases,
all based on the simple Harmonic oscillator. We feel that this
opens the way for a systematic investigation of the Berry phase
for one dimensional quantum systems.

This work has been partially supported by the European FEDER and by
the Spanish MCYT (Grant BFM2002-03773), MECD (Grant SAB2000-0240) and Junta de Castilla y
Le\'on (VA085/02). MLG thanks the Universidad de Valladolid for hospitality and support and
the NSF(USA) for partial support (Grant DMR-0121146).

%%%%%%%%%%%%%%%%%%%%%%%%%%%%%%%%%%%%%%%%%%%%%%%%%%%%%%%%%%%%%%%%%%%%%%%%%

\end{document}